\documentclass[10pt]{iopart}

\usepackage{iopams} 
\usepackage{graphicx}

\usepackage{cite}
\usepackage{hyperref}

\usepackage{color}
\usepackage{soul}
\setstcolor{red}

\begin{document}

\title[]{Zero-frequency supercurrent susceptibility signatures of trivial and topological zero-energy states in nanowire junctions}

\author{Lucas Baldo$^1$$^*$, Luis G. G. V. Dias Da Silva$^2$, Annica M. Black-Schaffer$^1$, Jorge Cayao$^1$$^\dagger$}

\address{$^1$ Department of Physics and Astronomy, Uppsala University, Box 516, S-751 20 Uppsala, Sweden}
\address{$^2$ Instituto de F\'{\i}sica, Universidade de S\~ao Paulo, Rua do Mat\~ao 1371, S\~ao Paulo, SP 05508-090, Brazil}
\ead{$^*$lucas.casa@physics.uu.se, $^\dagger$jorge.cayao@physics.uu.se}

\begin{indented}
\item[]Sep 2022
\end{indented}

\begin{abstract}
We propose a method to distinguish between trivial and topological, Majorana, zero-energy states in both short and long superconductor-normal-superconductor junctions based on Rashba nanowires using phase-biased equilibrium transport measurements. In particular, we show how the sawtooth profile of the supercurrent, due to the Majorana oscillation suppression in the topological phase for sufficiently long superconductor regions, leads to a strong signal in its zero-frequency susceptibility for a phase difference of $\phi=\pi$. This signal is notably insensitive to the chemical potential in the normal region, while trivial zero-energy states only causes signals in the susceptibility that is highly varying with the chemical potential, thus turning gating of the normal region into a simple experimental control knob.
Furthermore, we obtain that, by tuning the junction transparency, critical currents in both short and long junctions undergo a reduction in the number of oscillations as a function of magnetic field only in the topological phase, an effect that find to be intimately linked to Majorana non-locality.  Finally, we show that our results also hold at finite temperatures, thus highlighting their potential measurability under realistic experimental conditions.
\end{abstract}

\ioptwocol

\section{Introduction}
\label{section0}%

Majorana bound states (MBSs) have drawn much attention due to their non-locality and non-Abelian statistics \cite{Flensberg2021, Lutchyn2018}, which holds prospects for application in  topological quantum computation platforms.  MBSs have been shown to emerge in topological superconductors, a topological state of matter that can e.g.~be realized  by proximity inducing conventional $s$-wave superconductivity in semiconducting nanowires with Rashba spin-orbit coupling (SOC)\cite{PhysRevLett.105.077001, PhysRevLett.105.177002}.  In particular, in these nanowires, the topological phase is achieved by driving an external Zeeman field  above a critical value, $B_c$, after which MBSs appear as topologically protected zero-energy states at the edges of the topological regions. 

An important signature of MBSs is the zero-bias conductance peak (ZBCP) of height $2e^{2}/h$, theoretically predicted in NS junctions \cite{PhysRevLett.98.237002, PhysRevLett.103.237001, PhysRevB.82.180516} and at least partially consistent with existing experiments \cite{doi:10.1126/science.1222360, Higginbotham2015, doi:10.1126/science.aaf3961, Albrecht2016, Zhang2017, PhysRevLett.119.176805, PhysRevLett.119.136803, Gul2018}. The main disagreement stems from the fact that quantized ZBCPs can also be obtained from topologically trivial zero-energy Andreev bound states (ABSs) \cite{PhysRevB.86.100503, PhysRevB.86.180503, PhysRevB.91.024514, San-Jose2016, PhysRevB.96.075161, PhysRevB.96.195430, PhysRevB.97.155425, PhysRevB.98.245407, PhysRevLett.123.107703, PhysRevB.100.155429, PhysRevLett.123.217003, 10.21468/SciPostPhys.7.5.061, Avila2019, PhysRevResearch.2.013377, PhysRevLett.125.017701, PhysRevLett.125.116803, PhysRevB.102.245431, Yu2021, Prada2020, Pal2018, doi:10.1126/science.abf1513, PhysRevB.101.195303, PhysRevB.98.155314, PhysRevB.97.161401, PhysRevB.101.014512, PhysRevB.104.134507, PhysRevB.104.L020501, Marra_2022, PhysRevB.105.035148, Schuray2020, PhysRevB.102.045111, Grabsch2020, PhysRevB.102.245403, PhysRevB.103.144502,chen2022topologically,PhysRevB.106.014522}, implying that most of the reported experiments do not necessarily guarantee the existence of MBSs. These trivial zero-energy states (ZESs) can appear for fields well below $B_c$  in superconductor-normal-superconductor (SNS) junctions simply due to unavoidable chemical potential inhomogeneities and can display similar properties to MBSs \cite{Prada2020}. While in SNS junctions with short N regions, trivial ZESs might accidentally form and thus also be easily removed, in junctions with long N regions, they can also emerge robust against changes in the system parameters \cite{Prada2020}. In the long junction case, confinement and helicity play a crucial role for the formation of stable but trivial ZESs \cite{PhysRevB.91.024514}, where the helical regime is achieved when the chemical potential in the N region, $\mu_N$, lies within the helical gap opened by the SOC and Zeeman field $B$. Thus, the Zeeman field necessary to reach this helical phase often occurs below $B_c$ and can easily be mistaken for the topological phase transition at $B_c$ \cite{PhysRevB.104.L020501}. Unambiguous experimental signatures distinguishing MBSs and trivial ZESs are therefore still necessary. 

A promising approach toward the detection of topological zero-energy states is using equilibrium phase-biased transport, see e.g., Refs.\,\cite{PhysRevLett.112.137001, PhysRevB.96.024516,Tiira2017, PhysRevB.96.205425, cayao2018andreev,cayao2018finite,PhysRevB.94.085409,PhysRevLett.124.226801,PhysRevLett.126.036802,PhysRevB.100.035403, PhysRevB.95.155449, PhysRevB.94.205125, Ilan_2014,PhysRevB.106.L100502}. Here, due to a phase difference between two superconductors, a supercurrent flows across the junction, producing the Josephson effect \cite{Josephson1962}. The phase difference can be controlled by a magnetic flux in a SQUID geometry, which enables the determination of the phase-dependent supercurrent profile by sweeping the magnetic flux. In particular, it has been shown that, in junctions of nanowires with Rashba SOC, phase-dependent supercurrents in the topological phase are characterized by a discontinuity at $\phi=\pi$
in the limit of very long S regions, thus producing a characteristic sawtooth  profile as a clear sign of the topological phase \cite{PhysRevB.96.205425,Cayao_2018}. This property was partially used when phase-biased supercurrents were very recently predicted to be a useful tool for distinguishing between  MBSs and trivial ZESs \cite{PhysRevLett.123.117001,PhysRevB.104.L020501}, but this approach is still in its infancy and requires further analysis.

In this work, we exploit the phase-biased equilibrium transport and investigate how the zero-frequency supercurrent susceptibility is affected by the sawtooth profile developed in the topological phase in both short and long SNS nanowire junctions. We show that the zero-frequency supercurrent susceptibility develops distinct features that can be clearly related to the emergence of trivial ZESs and MBSs, respectively. In particular, we demonstrate that the zero-frequency supercurrent susceptibility is very sensitive to the chemical potential in the N region, $\mu_N$, when trivial ZESs are present but notably not when MBSs emerge, thus allowing for identification of the topological phase transition by simply gating the N region. We are able to explain this behavior by finding that the onset field for trivial ZESs depends on $\mu_N$, while $B_c$ does not, since the topological phase and its MBSs emerge in the S regions. This allows the zero-frequency supercurrent susceptibility to provide an easily accessible signature of the topological phase transition and thus become a tool for distinguishing trivial ZESs and MBSs.

Having identified the topological phase transition, we also exploit the MBSs energy oscillations as a function of magnetic field to test the spatial non-locality in the topological phase. In particular, we find a reduction in the number of oscillations of the zero-frequency supercurrent susceptibility when tuning the junction transmission, which allows us to identify the topological phase itself, beyond only finding the topological phase transition point. This is similar to what has been observed with critical currents in Ref.~\cite{cayao17b}, but here we go one step further and show that this effect also appears in the zero-frequency supercurrent susceptibility for both short and long junctions.  With the supercurrent susceptibility being experimentally accessible and already studied both theoretically and experimentally in similar systems \cite{PhysRevLett.122.076802, PhysRevLett.110.217001, Dassonneville_2018, PhysRevB.97.184505, PhysRevB.88.174505, PhysRevB.97.041415,PhysRevB.104.L100506}, our results should be within experimental reach. 

The remaining of this work is organized in the following way. In section~\ref{section1} we detail how we model the system, while in section \ref{discussionAndResults} we present and discuss our results. In particular, in~\ref{section2} we study the low-energy spectrum and explore the formation of trivial and topological ZESs in both short and long junctions. In~\ref{section3} we then use the energy spectrum to investigate and understand the supercurrent and its zero-frequency susceptibility in order to find signatures of ZESs and identify the topological phase transition. In~\ref{section4} we study the effects of lowering the junction transparency and how it can be used to identify the topological phase itself. Then, in ~\ref{section5} we establish the robustness of our results at finite but low temperatures. Finally, in section~\ref{sectionx} we summarize our results.

\section{Method}
\label{section1}
\begin{figure}[t!]
	\centering
	\includegraphics[width=\columnwidth]{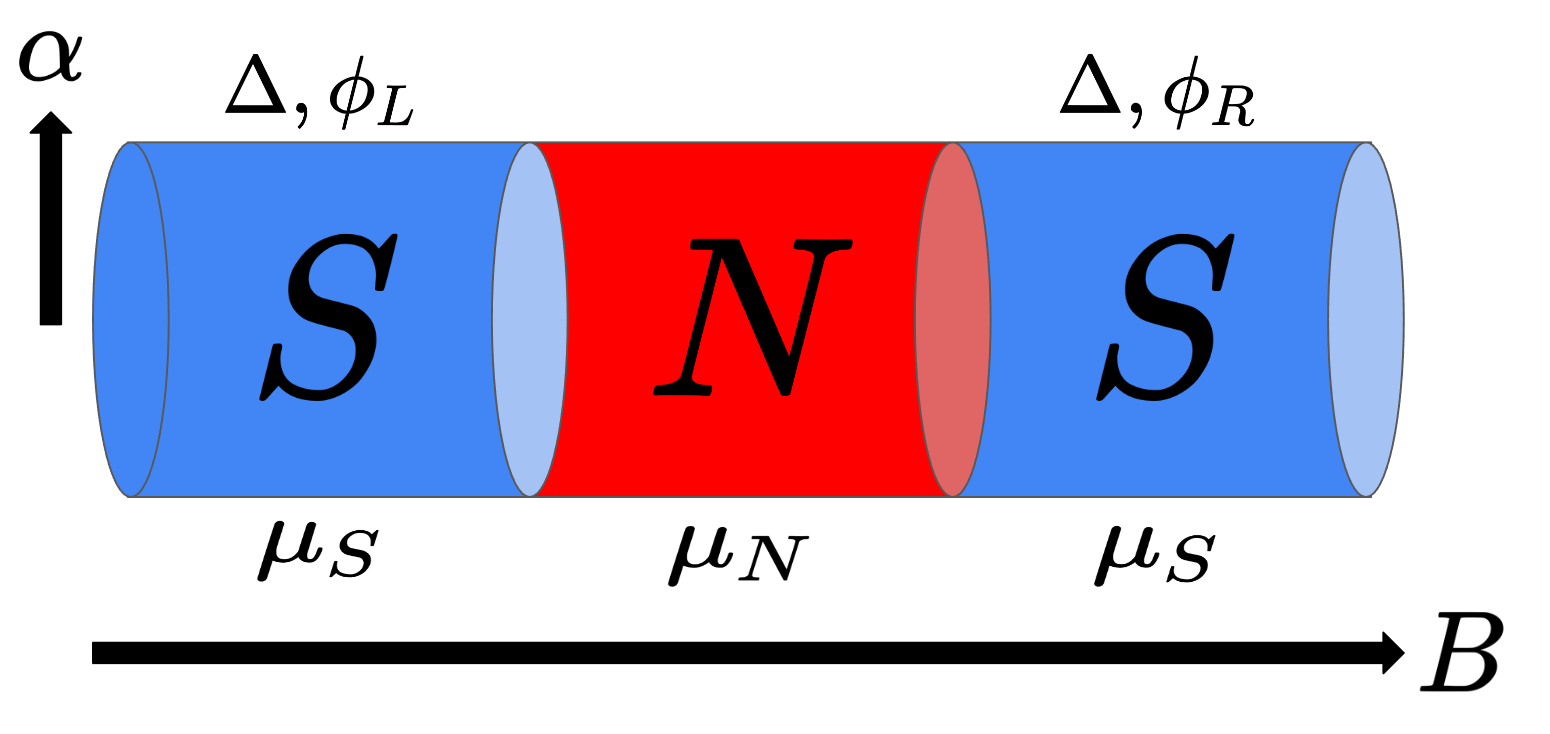}
	\caption{\label{fig:junction-diagram}Schematic diagram of an SNS junction. The junction is comprised of a single nanowire divided into three parts. In the outer parts (S regions) the order parameter $\Delta$ is taken to be finite, with its phase given by $\phi_{L/R}$, and the chemical potential is $\mu_S$. In the N region, the order parameter is zero and the chemical potential $\mu_N$. An external magnetic field $B$ is applied perpendicularly to the axis of the spin-orbit coupling of strength $\alpha$.}
\end{figure}

We consider SNS junctions made out of a single nanowire with Rashba spin-orbit coupling (SOC) of which two superconducting regions (S) are separated by a normal metallic region (N), as shown schematically in Fig.~\ref{fig:junction-diagram}. Superconductivity in the S regions is assumed to be proximity-induced into the nanowire by proximity to conventional $s$-wave superconductors. A finite controllable superconducting phase difference $\phi$ is allowed between the two S regions, thus creating a Josephson junction.

We model the nanowire as a single-channel one-dimensional system with the following effective Hamiltonian \cite{PhysRevLett.105.077001, PhysRevLett.105.177002}:
\begin{equation}
	H = \left( \frac{p_x^2}{2m} - \mu(x) \right) \tau_z + \frac{\alpha_R}{\hbar} p_x \sigma_y \tau_z + B \sigma_x \tau_z + \Delta(x) \sigma_y \tau_y, 
	\label{eq:Hamiltonian}
\end{equation}
where $p_x$ is the electron momentum along the nanowire, $m$ its effective electron mass, $\alpha_R$ is the Rashba SOC strength and $B$ is a Zeeman field, which comes from an external magnetic field applied perpendicularly to the spin-orbit axis. The Pauli matrices $\sigma_i$ and $\tau_i$ act on spin and particle-hole spaces, respectively. The spatially dependent chemical potential $\mu$ and superconducting order parameter $\Delta$ are defined piece-wise by
\begin{equation}
	\mu(x)=\cases{
	\mu_S &for $x \in S_L, S_R $ \\ 
	\mu_N &for $x \in N $\\}
\end{equation}
and
\begin{equation}
	\Delta(x)=\cases{
		\Delta e^{- i \phi/2} &for $x \in S_L$ \\
		0 &for $x \in N$ \\
		\Delta e^{+ i \phi/2} &for $x \in S_R$\\},
\end{equation}
where $S_L$ and $S_R$ represent the left and right S regions, respectively. Here, $\Delta$ corresponds to the proximity-induced superconducting order parameter in the nanowire.

We discretize the continuum model of Eq.~(\ref{eq:Hamiltonian}) into a tight-binding lattice with lattice constant $a=10$~nm and set $t = 25.4$~meV to be the nearest neighbor hopping, which corresponds to an effective mass $m = 0.015 m_e $, in accordance with experimental values for InSb \cite{Lutchyn2018, Mazur2022}. With this, we then model a SNS junction with N (S) regions of finite length $L_{N (S)}$, whose values are defined when presenting the respective junctions below. In order to control the junction transparency, we parametrize the hopping parameter at the NS interfaces, as $t_{NS}=\tau t$, where $\tau\ll1$ characterizes the tunnel regime and $\tau=1$ the full transparent regime.  Experimentally this can be achieved through gate-generated potential barriers at the NS interfaces.  Throughout this work, we use $\Delta = 0.25 $~meV and $\alpha_R = 20$\,meV\,nm, compatible with experimental values \cite{Lutchyn2018} and assume $\mu_N$ is highly variable and controlled by external gates.

Furthermore, we stress that here we   focus on  finite length SNS junctions with both short and long N regions. The terms short and long are not randomly chosen, but depends on the comparison between the superconducting coherence length $\xi$ and the length of the N region \cite{Beenakker:92}, where $\xi = \hbar v_F / (\Delta \pi)$ with $v_{F}$ being the Fermi velocity \cite{zagoskin}.  Thus,  short junctions are defined by $L_N < \xi$,  while long junctions by $L_N > \xi$.  For the parameters above we estimate $\xi$ to be between $\xi \sim 160-200$~nm for Zeeman field values between zero and $2 B_c$. Here we present results for short junctions with $L_N = 40 $~nm, while $L_N = 2000$~nm for long junctions. 

We are interested in trivial and topological ZESs in both short and long junctions. As discussed in the introduction, a very common mechanism for trivial ZESs, usually present in experiments, is the presence of chemical potential inhomogeneities, e.g., by considering the chemical potential in N and S to be different \cite{PhysRevB.91.024514,Prada2020}.  This is the regime we will consider here and under these conditions we will explore first the low-energy spectrum, and then the supercurrent and its zero-frequency susceptibility.

\section{Results and discussion}
\label{discussionAndResults}
In this part we present the low-energy spectrum and supercurrent and its susceptibility, obtained by using the model introduced in previous section. Before going further, it is perhaps important to point out the main features of Josephson junctions with MBSs when the chemical potentials in the N and S are the same, such that in the following sections we contrast when they are distinct and ZESs emerge. First, the model in Eq.~(\ref{eq:Hamiltonian}) develops a topological phase transition at  $B_c = \sqrt{\mu_S^2 + \Delta^2}$, above which MBSs emerge at the edges of the S regions \cite{PhysRevLett.105.077001, PhysRevLett.105.177002}. In a Josephson junction, the S regions thus become topological for $B>B_c$ and MBSs emerge at the end points of the S regions. At $\phi=0$ a MBS forms at each end of the outer sides of the SNS junction, but the two S region ends facing the N region do not host MBSs as they hybridize across the N region. However, at $\phi=\pi$ this hybridization becomes forbidden and two additional  MBSs emerge located at these inner edges of the SNS junction, at the interfaces between N and S regions \cite{PhysRevLett.108.257001,PhysRevLett.112.137001,PhysRevB.96.205425,cayao2018andreev,cayao2018finite}. For obvious reasons, we call MBSs at $\phi=0$ \textit{outer MBSs}, while we call the two additional MBSs at $\phi=\pi$ \textit{inner MBSs}. Below we explore signatures of both these MBSs and trivial ZESs that enable their distinguishability.

\subsection{Low-energy spectrum}
\label{section2}

\begin{figure}[t!]
	\centering
	\includegraphics[width=\columnwidth]{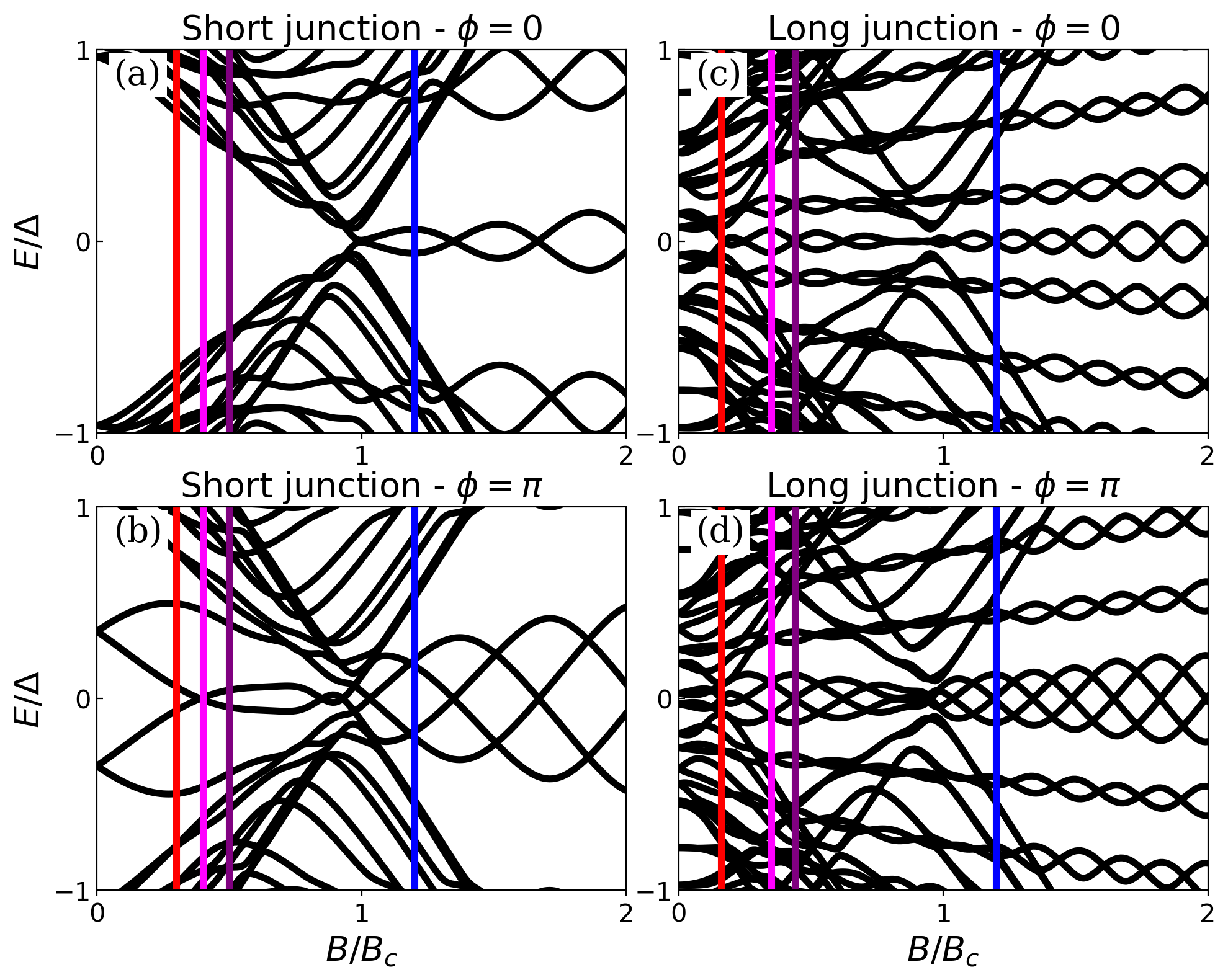}
	\caption{\label{fig:field-dispersion}Low-energy spectrum of short (a, b) and long (c, d) junctions as a function of Zeeman field $B$. (a, b) show the spectrum for a short junction with $\phi=0,~\pi$, respectively. (c, d) show the same for a long junction, where robust trivial ZESs appear, for both zero and finite $\phi$. Here $\mu_S=0.5$~meV, $\mu_N = 0.1$~meV, $L_S = 1000$~nm.}
\end{figure}

\begin{figure}[t!]
	\centering
	\includegraphics[width=\columnwidth]{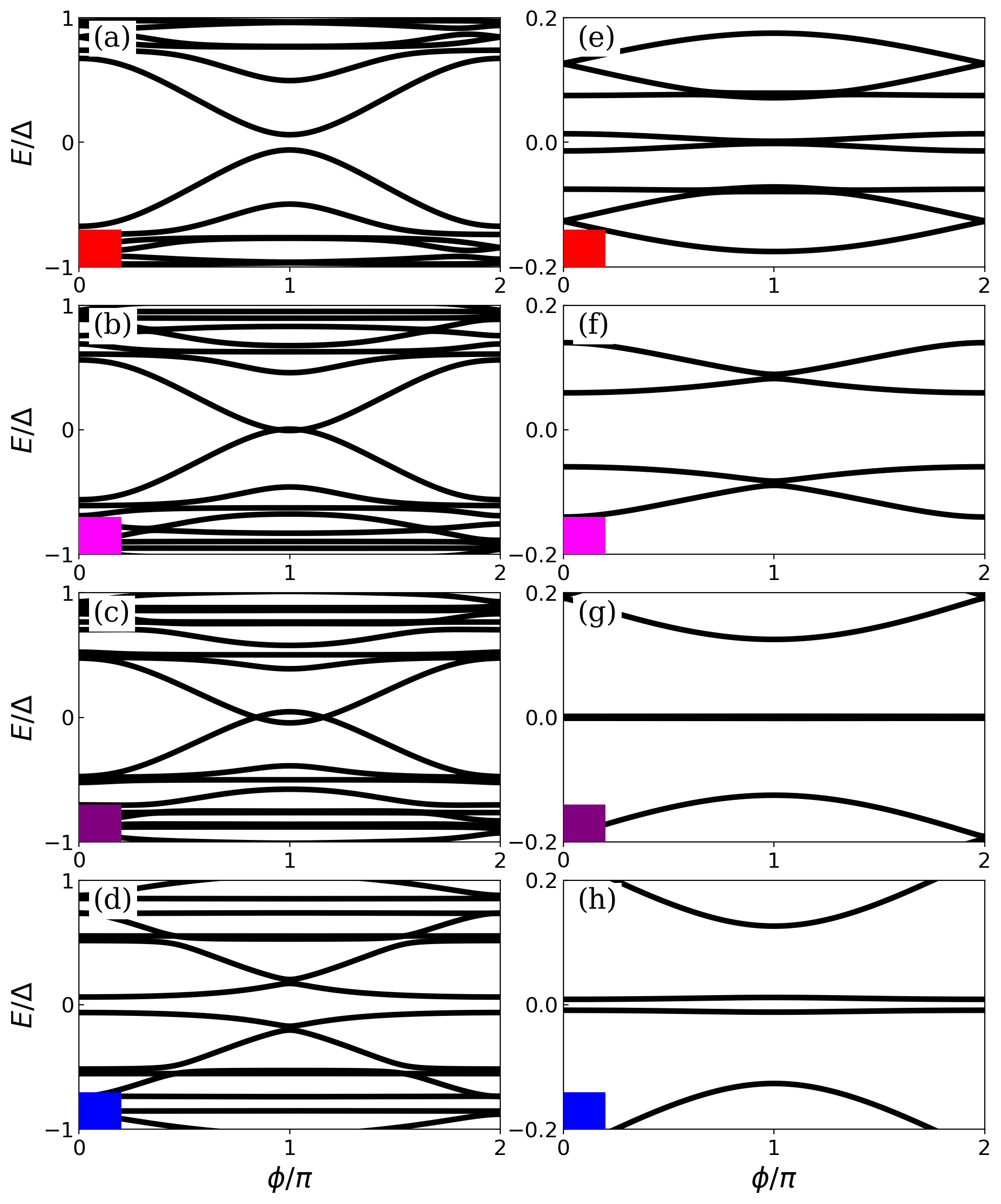}
	\caption{\label{fig:phase-dispersion} Low-energy spectrum of short (a, b, c, d) and long (e, f, g, h) junctions as a function of superconducting phase difference $\phi$. (a, b, c) show the spectrum for a short junction for magnetic fields around the emergence of trivial ZESs, where it is clear they emerge due to ABS crossings. (e) shows a similar field cut for a long junction at the first zero-energy crossing. (f, g) show the spectrum at a maximum energy split and a later zero-energy crossing, respectively. (d, h) show the spectrum in the topological regime. All field cuts are designated by colored lines in Fig.~\ref{fig:field-dispersion}(a-d), where the field values are $B/B_c = 0.3, 0.4, 0.5, 1.2$ for short junctions and $B/B_c = 0.16, 0.35, 0.44, 1.2$ for long junctions. Other parameters same as in Fig.~\ref{fig:field-dispersion}.}
\end{figure}

We start by presenting the low-energy spectrum based on the model described in section~\ref{section2} and identify the main features related to each type of ZESs, both of which can present similar experimental signatures, especially zero-bias conductance peaks. Later, we use the energy spectrum to calculate quantities that can distinguish between trivial and topological ZESs.

Fig.~\ref{fig:field-dispersion} shows the low-energy spectrum for both short (a, b) and long (c, d) junctions, and for phase values fixed at $\phi \!=\! 0$ (a, c) and $\phi=\pi$ (b, d) as a function of Zeeman field $B$. We here set $\mu_N$ such that the system hosts trivial ZESs. For both types of junctions, the system in the topological phase hosts two MBSs at $\phi=0$ (a, c)  while four MBSs at $\phi=\pi$ (b, d), appearing past a critical field $B_c$, in a similar way as when the chemical potentials in S and N are the same, see beginning of this section. Interestingly, for both short and long junctions, we observe the formation of trivial ZESs for fields well below $B_c$, evidenced by zero-energy crossings at low $B$. However, in short junctions these crossings usually appear only around $\phi=\pi$, whereas long junctions host them at $\phi=0$ as well \cite{PhysRevB.91.024514,PhysRevB.104.L020501}. The number of trivial ZESs varies with the value of $\mu_N$ and the length of the N region, with long junctions hosting multiple trivial ZESs and zero-energy crossings. This is due to confinement, since the N region is a metallic region between two superconductors that can host subgap energy states, which is also influenced by 
helicity effects in long junctions \cite{PhysRevB.91.024514,PhysRevB.104.L020501}. In addition, while the trivial ZESs of short junctions show a more accidental nature, evidenced by single crossings, long junctions can also host trivial ZESs that are robust for a range of parameter values. In particular, they can show robustness with respect to the Zeeman field, with multiple crossings before $B_c$, and also with respect to the superconducting phase difference.

By looking at the phase dispersion of these low-energy states we gain more insight into how trivial ZESs appear. In Fig.~\ref{fig:phase-dispersion}(a-c) we present the low-energy spectrum as a function of phase difference for a short junction with fixed values of the field just before, at, and after the trivial zero energy crossing marked in Fig.~\ref{fig:field-dispersion}(b) by vertical, color-coded lines. We see the zero-energy crossings occur as the Andreev bound states of the junction from both positive and negative energy sectors shift in energy as the field increases until eventually they touch at $\phi=\pi$ and cross each other. As an important consequence, this transition causes a change in the sign of the curvature of the energy-phase dispersion of the  lowest  energy state at $\phi=\pi$. As we will later show, this leaves a signature in the zero-frequency supercurrent susceptibility of the junction.

We have checked that, for a fixed value of the Zeeman field $B$, a change in $\mu_N$ either moves the phase at which the short junction trivial ZESs appear away from $\phi=\pi$ or removes them altogether. This is in stark contrast with the behavior of the topological MBSs, for which we show in Fig.~\ref{fig:phase-dispersion}(d) the phase-dependent spectrum in a short junction. The almost dispersion-less states close to zero energy are the MBSs localized at the outer edges of the junction. 
Close to $\phi \!=\! \pi$, another pair of states also approaches zero energy, corresponding to the MBSs of the inner edges of the S regions, and eventually hybridizes with the outer pair. The hybridization energy oscillates with Zeeman field, but decays when increasing $L_S$. In Figs.~\ref{fig:field-dispersion} and \ref{fig:phase-dispersion} we use $L_S = 1000$~nm, such that this energy split is visible at $\phi=\pi$. As $L_S$ increases, the split diminishes, becoming negligible for $L_S \gtrsim 3000$~nm. This splitting is part of an anti-crossing structure and for longer $L_S$ the dispersion of the inner MBSs gains a linear character.

In order to describe the low-energy spectrum further, it is instructive to consider the limit of semi-infinitely long S regions. In this limit and in the topological regime, the outer MBSs are not present and the low-energy spectrum is dominated by the inner MBSs. Their phase dispersion $\epsilon^{(S)}_{\pm} (\phi)$ in a short junction is then described by \cite{kwon2004fractional, PhysRevB.79.161408}
\begin{equation}
	\epsilon^{(S)}_{\pm} (\phi) = \pm \epsilon^{(S)}_{0} \vert \cos (\phi/2) \vert,
	\label{eqn:short-dispersion}
\end{equation}
where $\epsilon^{(S)}_{0}$ depends on system parameters such as SOC strength, Zeeman field, and junction transparency. For a topological junction with perfect transmission, the proportionality constant takes the value of the induced gap $\epsilon^{(S)}_{0} = \Delta_\textrm{ind}$, but in general this state is a subgap state with $\epsilon^{(S)}_{0} < \Delta_\textrm{ind}$. Since we are interested in the low-energy behavior of the state around $\phi=\pi$, it is useful to expand the dispersion around this point and we find, up to second order,
\begin{equation}
	\epsilon^{(S)}_{\pm} (\phi) \approx \pm \frac{1}{2} \epsilon^{(S)}_{0} \vert \phi - \pi \vert.
	\label{eqn:short-dispersion-approx}
\end{equation}
As seen in Fig.~\ref{fig:phase-dispersion}(d), in a system with finite S regions this linear dispersion gives way to an anti-crossing structure. By increasing the length of the S regions the anti-crossing energy splitting becomes suppressed and the dispersion becomes more linear around $\phi=\pi$. Since this splitting and the linearity of the dispersion depend on the length of S regions, they can be viewed as a direct signature of the non-locality of the topological phase and its MBSs.

Next, we turn to long junctions. In a long junction, the number of states introduced by the normal region greatly increases, making the field evolution of the phase-dispersion more complex. In the topological regime, we find that the low-energy spectrum shows an oscillating hybridization of the MBSs similar to a short junction, but with a different oscillation period, as can be seen e.g.~comparing Fig.~\ref{fig:field-dispersion}(a,b) to Fig.~\ref{fig:field-dispersion}(c,d). In Fig.~\ref{fig:phase-dispersion}(h) we show the phase dispersion in the topological regime at the same field value as for the short junction in Fig.~\ref{fig:phase-dispersion}(d). By comparing the panels we see that at this particular magnetic field the junctions have opposite hybridization behavior. While the short junction shows two states degenerate at $\phi=\pi$ at a slightly larger energy, the long junction shows two outer MBSs with a nearly flat dispersion. 

Another important difference  between the short and long junctions is that for long junctions in the trivial regime, although we do observe ABS crossings similar to short junctions [Fig.~\ref{fig:phase-dispersion}(e)], it is also common for the ABSs to completely shift above or below zero energy for all values of $\phi$, especially when increasing $B$. They can then oscillate between hybridizing with a higher energy state, such as in Fig.~\ref{fig:phase-dispersion}(f), or having a flat dispersion, such as in Fig.~\ref{fig:phase-dispersion}(g), which is similar to the behavior of the topological MBSs found in Fig.~\ref{fig:phase-dispersion}(d,h). At $\phi = 0$ this happens in the helical regime of the N region ($B > \mu_N$) \cite{PhysRevB.91.024514} and it generates robust trivial ZESs that experience multiple zero energy crossings in the trivial regime when tuning $B$. We also calculate the spectrum for different $\mu_N$ values and find that, as in the case of short junctions, the onset of these crossings is dependent on $\mu_N$, which enables differentiation from the topological MBSs.

Finally, we also consider the semi-infinite $L_S$ limit of a long junction in the topological regime, where the dispersion can be approximated by \cite{PhysRevLett.110.017003}
\begin{equation}
	\epsilon^{(L)}_{\pm} (\phi) = \pm \epsilon^{(L)}_{0} \vert \phi - \pi \vert.
	\label{eqn:long-dispersion}
\end{equation}
Around $\phi=\pi$ this behaves similarly to the case of short junctions. However, in the long junction case, a perfect transmission now leads to a proportionality constant $\epsilon^{(L)}_{0} = \xi \Delta/(2 L_N)$ \cite{PhysRevLett.110.017003}.
Even though the expressions in Eqs.~(\ref{eqn:short-dispersion-approx}) and (\ref{eqn:long-dispersion}) are for ideal  semi-infinite S regions, it helps understanding of the MBSs in our system which is of finite size and thus  more complicated. Indeed, we find that a main qualitative feature remains, which is that the dispersion becomes linear ($\epsilon_\pm^{(S/L)} \propto \pm \vert \phi - \pi \vert$) around $\phi=\pi$ also for only moderately large $L_S$.  

In summary, we find that, although the emergence of topological MBSs, always occurs above $B_c$, trivial ZESs can easily emerge before this transition. These states have different origins, as MBSs arise due to the topological phase, while trivial ZESs emerge due to ABS zero-energy crossings which strongly depend on the chemical potential of the normal region. Furthermore,  the behavior of MBSs at $\phi = \pi$ for both types of junctions induce a linear dispersion proportional to $  \pm \vert \phi - \pi \vert$. This causes the derivative of the dispersions to be discontinuous at $\phi=\pi$ and consequently the second derivative will diverge. It is this behavior that we will exploit in the next subsection.

\subsection{Supercurrent and zero-frequency susceptibility}
\label{section3}
In the previous subsection, we calculated the low-energy spectrum and showed that both short and long junctions can easily host accidental and robust trivial ZESs, respectively. These states present an obstacle in the detection of MBSs and methods of differentiating these trivial states from the topological MBSs are urgently needed. In order to propose a solution to this issue, we next investigate the supercurrent of the junction and also its susceptibility. The phase-dependent supercurrent of a Josephson junction can be directly obtained from its spectrum as \cite{Beenakker:92,PhysRevB.96.205425} 
\begin{equation}
	I(\phi) = -\frac{e}{\hbar} \sum_{\epsilon_p>0} \tanh\Big(\frac{\epsilon_p}{2 \kappa_{B} T}\Big) \frac{d \epsilon_p}{d\phi},
	\label{eq:supercurrent}
\end{equation}
where $\kappa_{B}$ is the Boltzmann constant, $T$ the temperature and the summation is performed over positive eigenvalues. Another quantity we are interest in is the supercurrent susceptibility \cite{PhysRevLett.122.076802,	PhysRevLett.110.217001,	Dassonneville_2018,	PhysRevB.97.184505,	PhysRevB.88.174505,	PhysRevB.97.041415}, and in particular, the \textit{zero-frequency supercurrent susceptibility} 
\begin{equation}
	S(\phi) = - \frac{d I }{d \phi}.
	\label{eq:susceptibility}
\end{equation}
As it will be obvious later, we are here mostly focusing on the value this quantity takes at $\phi=\pi$, which we denote simply by $S_\pi = S(\phi=\pi)$. We investigate how the above quantities depend on system parameters and begin by considering the zero temperature case, while finite temperature effects are covered in section~\ref{section5}.  

\subsubsection{Short junctions}
\label{sec:short-junction}

\begin{figure}[t!]
	\centering
	\includegraphics[width=\columnwidth]{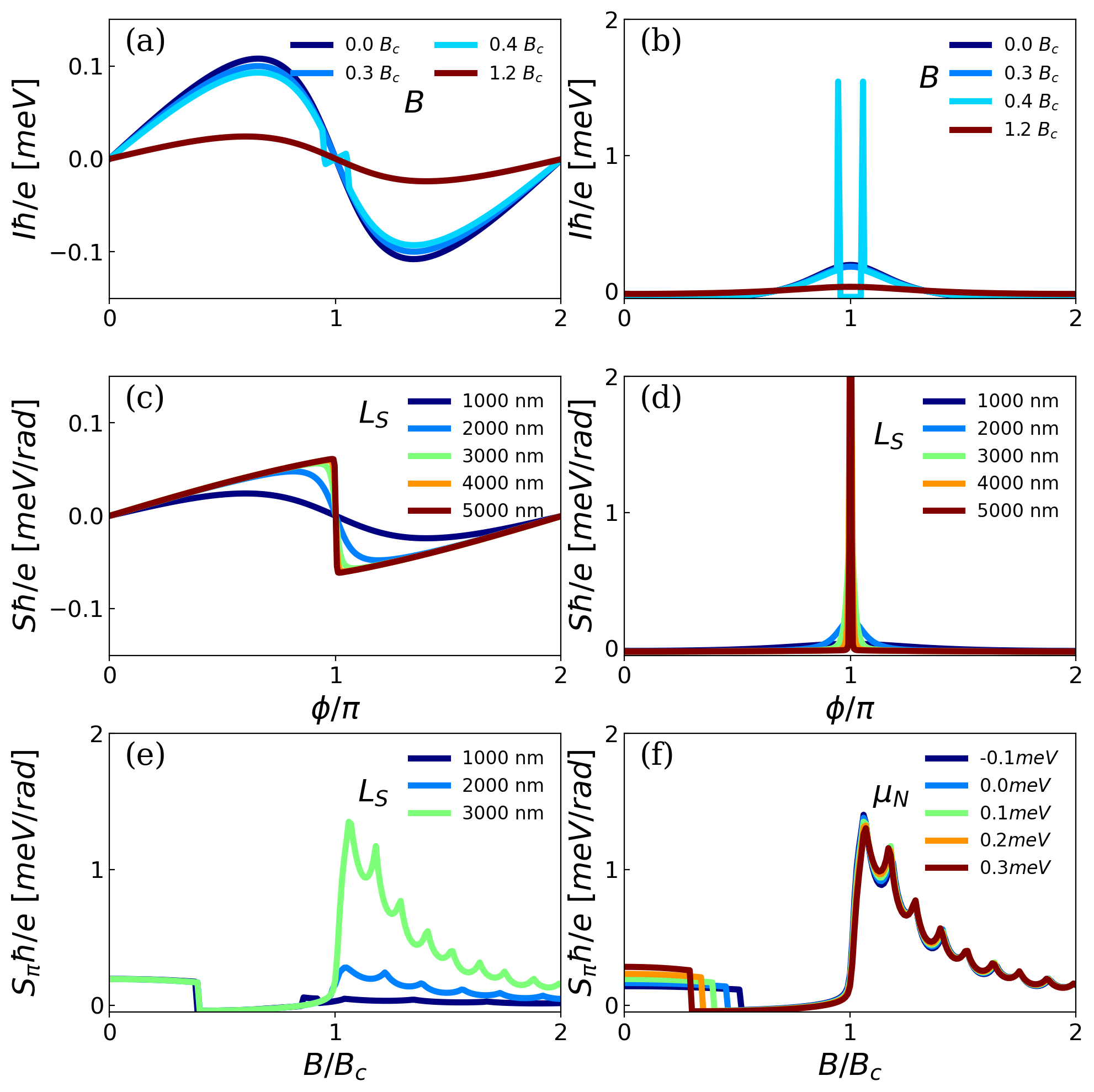}
	\caption{\label{fig:CS-short} Supercurrent and zero-frequency susceptibility for short junctions hosting trivial ZESs and MBSs. (a,c) Supercurrent $I$ as a function of phase difference $\phi$ for different values of the Zeeman field $B$ and S region lengths $L_S$, respectively. In (c) we show the topological regime at $B=1.2 B_c$. (b,d) Corresponding zero-frequency supercurrent susceptibility. (e,f) Zero-frequency supercurrent susceptibility at $\phi = \pi$, $S_\pi$, as a function of Zeeman field $B$ for different values of $L_S$ and $\mu_N$, respectively. Here $L_S = 1000$~nm in (a,b) and $L_S = 3000$~nm in (f), while $\mu_N = 0.1$~meV in (a-e). We also set $\mu_S=0.5$~meV. Supercurrent and susceptibilities are in units of $e/\hbar \textrm{meV}$ and $e/\hbar \textrm{meV}/\textrm{rad}$, respectively.}
\end{figure}

In Fig.~\ref{fig:CS-short} we show results for a short junction. In Fig.~\ref{fig:CS-short}(a) we plot the supercurrent as a function of the phase for different values of the Zeeman field, using $L_S = 1000$~nm, the same as in Figs.~\ref{fig:field-dispersion} and \ref{fig:phase-dispersion}. At low fields, the supercurrent shows the usual sinusoidal character \cite{Josephson1962}. At higher fields, but still in the trivial regime, the supercurrent develops a zigzag profile when trivial ZESs emerge. This zigzag profile begins as a pair of discontinuities near $\phi=\pi$, which then move away from each other as the field increases further, due to the ABS crossings moving away from $\phi=\pi$. Above $B_c$ the supercurrent reverts back to the sinusoidal profile, but with a smaller amplitude. This picture changes, however, if we consider junctions with longer S regions, as shown in Fig.~\ref{fig:CS-short}(c), where we show the supercurrent profile in the topological regime for several different values of $L_S$. 

In order to better understand the change in supercurrent profile induced by changing $L_S$, we first point towards the $\epsilon_\pm^{(S)} \propto \pm \vert \phi - \pi \vert$ dispersion of junctions with infinitely long S regions in the topological regime around $\phi=\pi$, see Eq.~(\ref{eqn:short-dispersion-approx}). This dispersion leads to the development of a sawtooth profile, with the discontinuity pinned at $\phi=\pi$. For any finite S regions, however, the energy dispersion is never perfectly linear, such that the sawtooth profile becomes smoothed out. We still find that the sawtooth profile is prominent for $L_S \gtrsim 3000$~nm, but gives way to a sine-like curve for values below that.

The supercurrent behavior  in Fig.~\ref{fig:CS-short}(a,c) have corresponding features in its susceptibility profile [Fig.~\ref{fig:CS-short}(b,d)], as they are related by a derivative, see  Eqs.\,(\ref{eq:supercurrent}) and (\ref{eq:susceptibility}).
In fact, discontinuities in the supercurrent lead to pronounce susceptibility peaks with important consequences. In the trivial regime, the zigzag profile in the supercurrent induced by the trivial ZESs leads to susceptibility peaks symmetrical around $\phi=\pi$, which can form further away from $\phi=\pi$ for a small variation of the system parameters. In the topological regime, the reduced supercurrent amplitude leads to a reduced susceptibility amplitude but the sawtooth profile developing for larger $L_S$ leads to clear susceptibility peaks pinned at $\phi=\pi$.
The height of these susceptibility peaks increases with $L_S$ and in the limit of infinitely long S regions they even diverge. With the underlying dispersion being a direct consequence of the MBSs and their separation across the S regions, the developing supercurrent susceptibility peak at $\phi=\pi$ is a consequence of the MBSs non-locality. The interesting dependence  on $L_S$ means that we can always design a longer system to improve the visibility of the susceptibility signal from the topological phase, without affecting the response of trivial ZESs.

Next, in Fig.~\ref{fig:CS-short}(e) we study in more detail how $S_\pi$ varies with the Zeeman field $B$ for different values of $L_S$. We see that at low fields, $S_\pi$ takes on a finite, approximately constant value. This is because in this regime the low-energy spectrum of the positive energy sector is dominated by two ABSs with positive curvatures. As the field is increased, the ABSs split in energy and the lowest-energy branch  eventually crosses zero, as seen in Fig.~\ref{fig:phase-dispersion}(a-c). This causes the lowest energy state in the positive energy sector to get an opposite curvature, which in turn leads to a sharp jump in the $S_\pi$ profile with an accompanying change of sign. As the field is increased further, the system eventually enters the topological regime, where MBSs emerge. For systems with short S regions, the MBSs from both edges of the same S region hybridize significantly, leading to energy oscillations with the Zeeman field, but this has very minor impact on the zero-frequency susceptibility at $\phi=\pi$. For sufficiently long S regions the inner MBSs acquire a linear dispersion around $\phi=\pi$, leading to the development of a strong peak signature in $S_\pi$. Since this is a consequence of the topological phase transition, this peak is pinned to $B_c$. Deeper into the topological regime $S_\pi$ decays with increasing $B$.

In order to further differentiate between signatures in $S_\pi$ coming from trivial or topological ZESs, we exploit the fact that the trivial ABS crossings are strongly influenced by $\mu_N$, the chemical potential in the N region, while $B_c$ depends only on $\mu_S$, as it is tied to the topological phase transition in the bulk S region. This means that tuning $\mu_N$ should only affect features the zero-frequency susceptibility coming from the emergence of trivial ZESs, and not the topological MBSs. In Fig.~\ref{fig:CS-short}(f) we plot $S_\pi$ as a function of Zeeman field for a fixed value of $L_S = 3000$~nm for varying values of $\mu_N$. The value of $L_S$ is chosen such that the junction display a sawtooth profile in the supercurrent due to the MBSs. We observe that the step-like features induced by trivial ZESs are present for all values of $\mu_N$, but notably the field values at which they occur always differ. On the other hand, the peak induced by the topological phase transition is always stuck at $B_c$. This means that by simply gating the N region we can directly and clearly identify which features in the $S_\pi$ profile have a topological nature and which ones do not.

\subsubsection{Long junctions}
\label{sec:long-junction}

\begin{figure}[t!]
	\centering
	\includegraphics[width=\columnwidth]{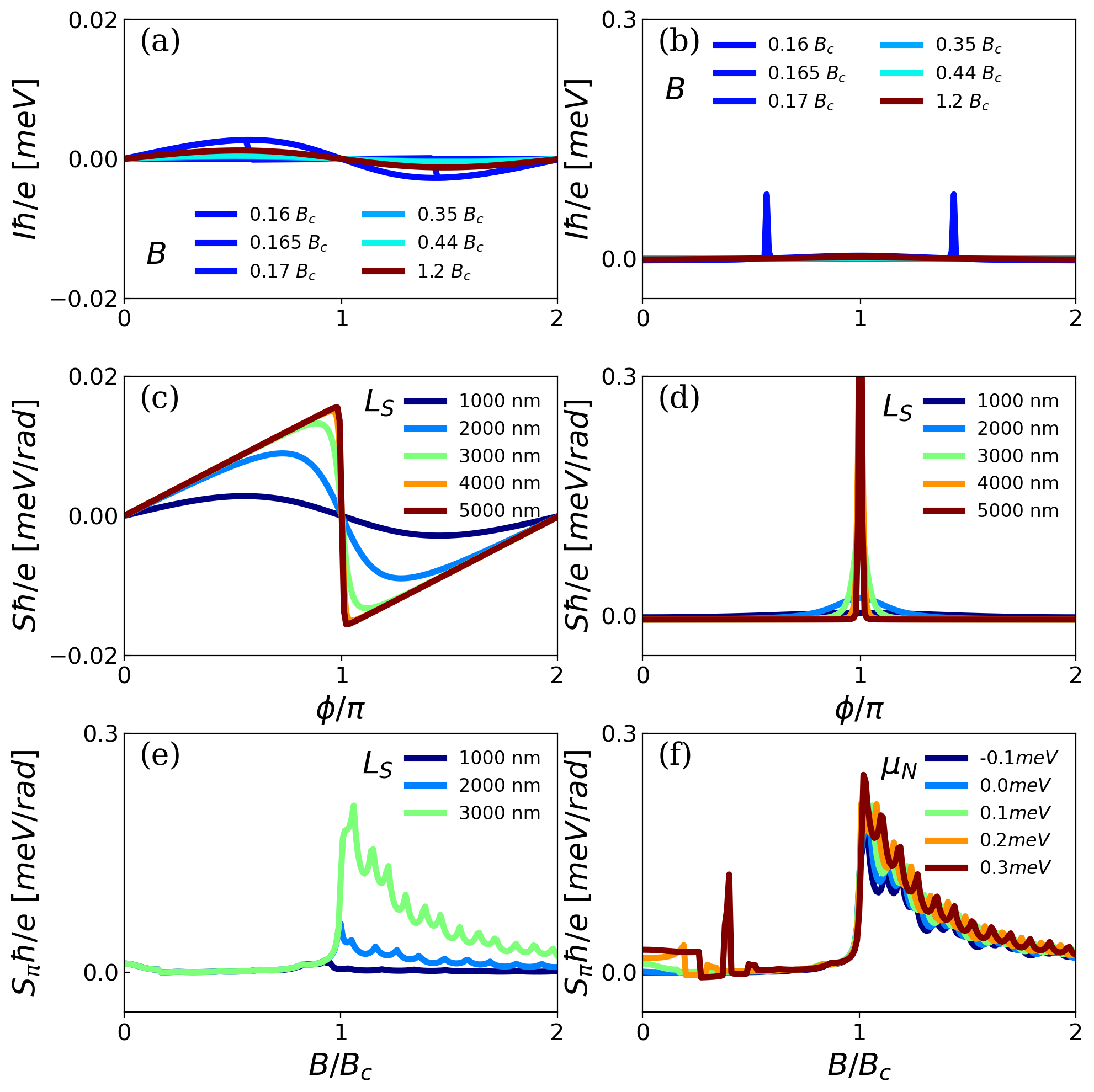}
	\caption{\label{fig:CS-long} Similar as Fig.~\ref{fig:CS-short}, but for long junctions. (a,c) and (b,d) show the supercurrent and its zero frequency susceptibility, respectively, as a function of the phase difference $\phi$ for different Zeeman fields $B$ and superconducting lengths $L_S$. (e,f) Zero-frequency supercurrent susceptibility at $\phi=\pi$, $S_\pi$, as a function of Zeeman field $B$ for different $L_S$ and $\mu_N$, respectively. Here same parameters as in Fig.~\ref{fig:CS-short} are used, except for $L_N = 2000$~nm and the field values in (a,b), indicated in the figure.}
\end{figure}

Next, we consider junctions with a long N region. Figure~\ref{fig:CS-long} shows numerical results for the same quantities depicted in Figure~\ref{fig:CS-short}, but for a long junction with $L_N = 2000$~nm. The first two panels, (a,b), show how the supercurrent and its zero-frequency susceptibility evolve for different values of the magnetic field, for the same chemical potential configuration as in Figure~\ref{fig:CS-short}. The trivial ZESs result from ABS crossings, bringing the emergence of a zigzag profile around this point in the supercurrent. However, the crossing lasts only for a short interval of $B$ values (within one percent of $B_c$), which means this zigzag feature and their corresponding peaks in susceptibility disappear quickly as $B$ is increased. By further increasing the field, many other trivial ZESs are formed as the lowest energy state oscillates between having no phase dispersion and hybridizing at $\phi=\pi$, as displayed in Fig.~\ref{fig:phase-dispersion}(f, g). As such, they do not contribute significantly to $S_\pi$.

In the topological regime, $S_\pi$ develops a strong dependence on $L_S$, as can be seen in Fig.~\ref{fig:CS-long}(c,d). Similarly to the case of short junctions, the origin of this dependence can be understood from the MBSs phase dispersion for infinitely long $L_S$, $\epsilon_\pm^{(L)} \propto \pm \vert \phi - \pi \vert$, which leads to a sawtooth supercurrent profile and also a diverging $S_\pi$. We illustrate this in  Fig.~\ref{fig:CS-long}(e) where we show how $S_\pi$ varies with the Zeeman field in a long junction with varying $L_S$ and overall we observe a behavior much similar to that of short junctions of Fig.~\ref{fig:CS-short}(e). 
Finally, in Fig.~\ref{fig:CS-long}(f) we fix $L_S=3000$~nm and show trivial ZESs features are highly dependent on $\mu_N$ in long junctions, but that peaks associated with the topological phase transition and MBSs are not sensitive to $\mu_N$, just as in short junctions. We note that there are more abrupt changes and a more irregular behavior for the trivial ZESs in long junctions, as compared to a short junction. This can be attributed to the more complicated energy spectrum in the former. Still, we observe a noticeable variation of the profile of $S_\pi$ in the trivial regime as a function of $\mu_N$, while the feature introduced by the topological phase transition remains unchanged.
This means that $S_\pi$, in conjunction with varying the gating in the N region, can be used to clearly identify the topological phase transition in both long and short junctions. 

For the parameters considered in this subsection, the short junction hosts phase-induced (finite $\phi$) trivial ZESs, and the long junction hosts both zero-phase ($\phi = 0$) and phase-induced trivial ZESs. We have also checked that our overall conclusions remain the same in chemical potential configurations where also zero-phase trivial ZESs are present in short junctions, such as for the effective values reported in Refs.~\cite{PhysRevLett.123.117001, PhysRevB.105.144509}. We also remark that our conclusions depend on the phase dispersion of the inner MBSs pair not being completely flat. This may happen when the chemical potential configuration is such that it does not allow hybridization of the inner MBSs at $\phi = 0$, leading to a vanishing proportionality factor in the long S region limit dispersion, $\epsilon_\pm^{(S/L)} \propto \pm \vert \phi - \pi \vert$. We avoid such pathological regimes in our calculations, but focus on parameter regions where $\vert \mu_N \vert < \vert \mu_S \vert$, where we never found such issues. In this work we also set $\mu_S = 0.5$~meV, but we checked that our conclusions hold for other values of $\mu_S$ for long enough S regions.
As a consequence, we find that gating the N region and measuring $S_\pi$ can be used as an easy, distinctive tool to differentiate trivial ZESs from MBSs appearing in the topological regime, in both short and long junctions.

\subsection{Transparency effects}
\label{section4}
In the previous subsection, we identified the topological phase transition and differentiated it from features produced by trivial ZESs in the zero-frequency supercurrent susceptibility measured at $\phi \!=\! \pi$. Here we show how changing the junction \emph{transparency}, which can be done by tuning localized gates at the NS interfaces, allows us to additionally determine if the junction is inside the topological phase or not, still using only measurements of $S_\pi$. As a consequence, $S_\pi$ can both determine the topological phase transition and the topological phase itself.
The latter is done by exploiting the energy oscillations of the MBSs as a function of Zeeman field. We note that a similar study has been done in Ref.~\cite{cayao17b}, where oscillations in the critical current of short junctions were found and then attributed to the MBSs. Here we investigate similar oscillations, but in $S_\pi$ instead in order to show how $S_\pi$ only can be utilized to unambiguously determine the topology. For this, we turn our attention to the $S_\pi$ oscillations observed for both short and long junctions at $B>B_c$ in the previous figures. We thus fix  $L_S=3000$ nm, which we already showed is enough for a clear signal of the topological phase transition in $S_\pi$.

\begin{figure}[t!]
	\centering
	\includegraphics[width=\columnwidth]{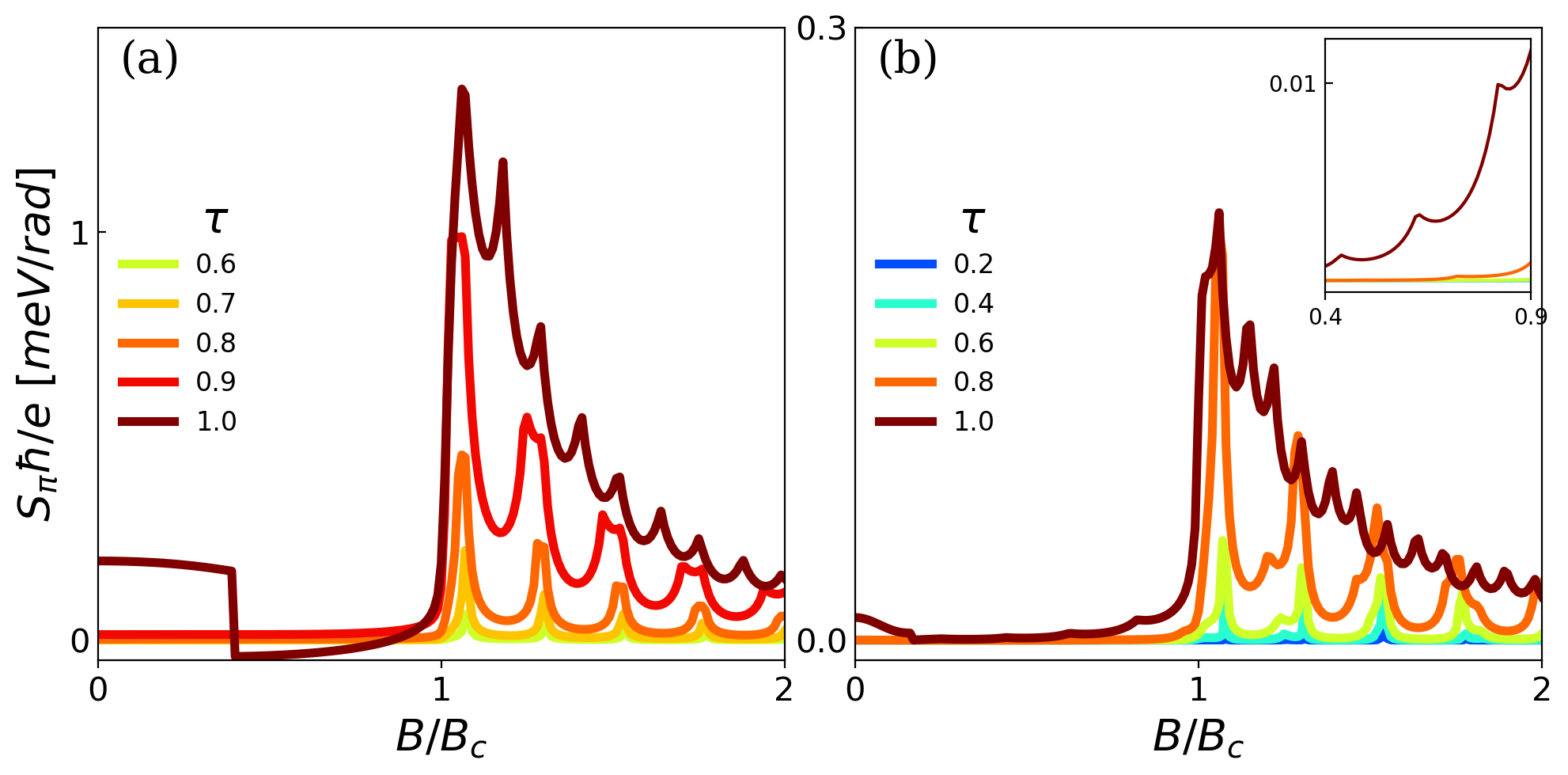}
	\caption{\label{fig:transparency}Transparency dependence of $S_\pi$. (a,b) show $S_\pi$ as a function of Zeeman field $B$ in the topological regime for different transparency values $\tau$ in a short and long junction, respectively. An approximate doubling of the oscillation period is observed as $\tau$ is reduced from $1$ to values in the tunneling regime. Inset in (b) shows $S_\pi$ oscillations in the trivial regime present in long junctions. Here $L_S = 3000$~nm, $\mu_N = 0.1$~meV and $\mu_S=0.5$~meV.}
\end{figure}

For a high transparency junction at $\phi \!=\! \pi$ in the topological regime, the two pairs of MBSs oscillate out-of-phase with respect to each other. This gives rise to periodic, intermittent crossings at zero energy as the pairs switch ordering as the pair closer or further apart from zero energy, see Fig.~\ref{fig:field-dispersion}(d). In short junctions, as the transparency is lowered, these energy oscillations have been observed to align such that the zero-energy crossings of MBSs happen at the same Zeeman fields \cite{cayao17b}. As a consequence, zero-energy crossings of MBSs in short junctions were shown to happen at approximately double the period compared to the high transparency case.  This period doubling effect is depicted in Fig.~\ref{fig:transparency}(a), where we let $\tau$ vary from the ideal transparency limit ($\tau = 1$) to the short junction tunneling regime ($\tau \sim 0.6$). For long junctions in Fig.~\ref{fig:transparency}(b), we observe a similar effect, but in this case, at the same lower value of $\tau \sim 0.6$, a crossover regime between higher and lower periodicity  is still found. In contrast to short junctions, however, for $\tau \sim 0.6$ an extra oscillation with small amplitude accompanies the pronounce peaks, leading to a deformation in the oscillatory pattern of $S_\pi$ at intermediate values of $\tau$.  This is a consequence of long junctions achieving the tunneling regime only at lower $\tau$ values. It is only by lowering the transparency to $\tau \sim 0.3$ that the long junction finally arrives at the tunneling regime and we are able observe a well-defined larger period oscillation, similar to the period doubling effect in short junctions [Fig.~\ref{fig:transparency}(a)].

We finally note that in long junctions the robust trivial ZESs can also show energy oscillations with the Zeeman field, which lead to corresponding oscillations in $S_\pi$. This is displayed in the inset in Fig.~\ref{fig:transparency}(b). However, these oscillations have a reduced amplitude and die out more quickly than their topological counterparts as $\tau$ decreases and hence they do not show the period doubling effect. Thus, we show that tuning the junction transparency can be used to identify the topological phase through a period doubling effect of MBSs oscillations as a function of magnetic field. 

\subsection{Temperature effects}
\label{section5}

\begin{figure}[t!]
	\centering
	\includegraphics[width=\columnwidth]{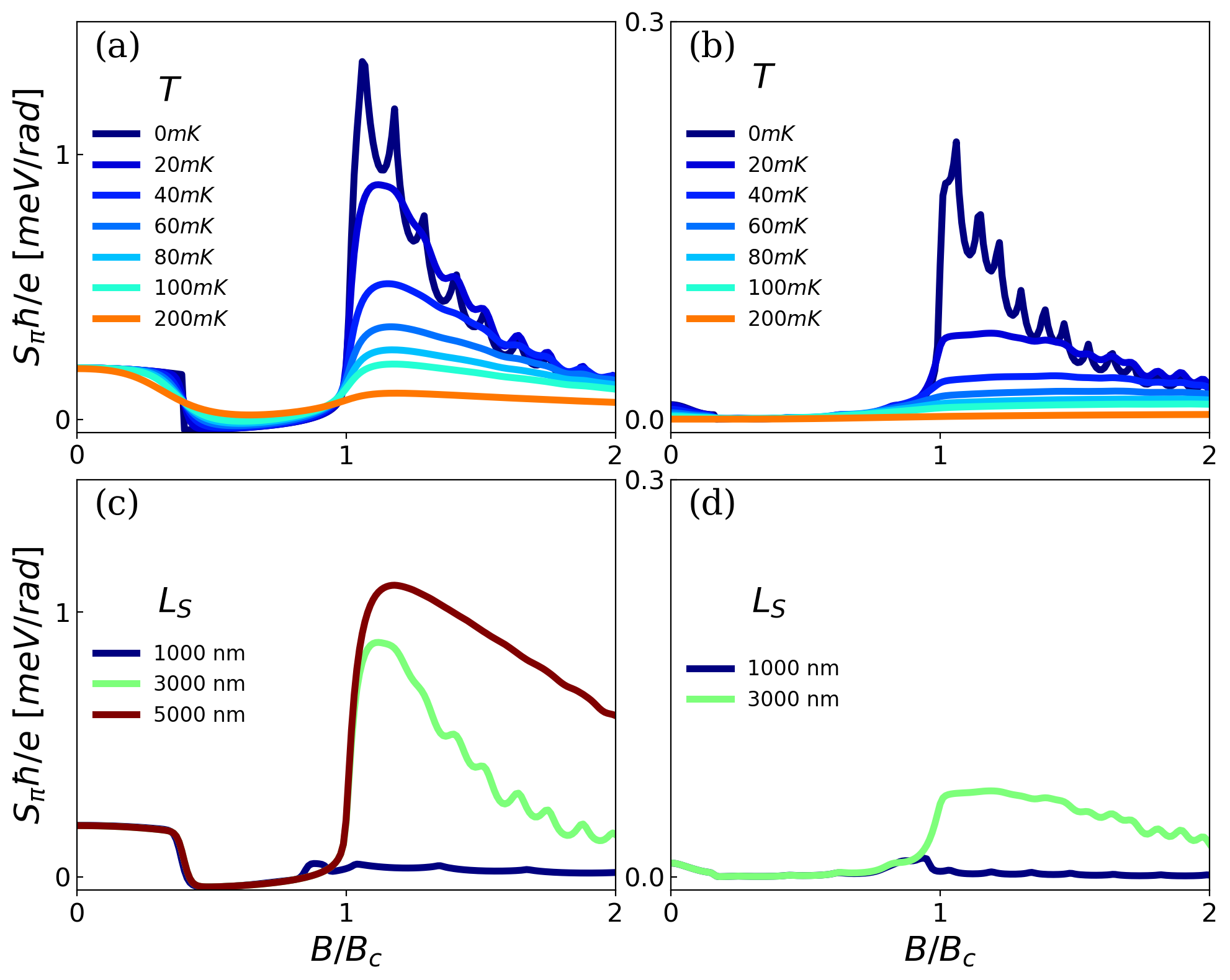}
	\caption{\label{fig:temperature}Temperature effects on $S_\pi$ for short (a, b) and long (c, d) junctions. (a, b) $S_\pi$ as a function of Zeeman field for different temperatures, for $L_S = 3000$~nm. (c, d) $S_\pi$ as a function of Zeeman field for different values of $L_S$ and a fixed temperature $T=20$~mK. We use $\mu_S=0.5$~meV.}
\end{figure}

The results we have presented so far were all calculated in the zero-temperature limit. In order to showcase their experimental relevance, we now consider finite temperature effects. As a direct consequence of the hyperbolic tangent term in Eq.~(\ref{eq:supercurrent}), signatures in the supercurrent coming from low-energy states are suppressed much more strongly for energies below $2 k_B T$. This means that at higher temperatures, near-zero-energy states have their contribution to $S_\pi$ strongly suppressed. In particular, the sawtooth profile in the supercurrent  is affected by this suppression, as well as signatures coming from trivial ZESs. 
Figure~\ref{fig:temperature}(a,b) show how the effects of a finite temperature also strongly affect $S_\pi$ in short and long junctions, respectively. We observe that, for both short and long junctions, at a temperature of a few hundred mK, $S_\pi$ takes values much smaller than its zero temperature value in the topological regime. For concreteness, at $T\sim200$~mK we observe that in short junctions the highest signal of $S_\pi$ in the topological regime becomes more than one order of magnitude smaller than its zero temperature value. In long junctions the same ratio is over 50.
Interestingly, in short junctions the features introduced by trivial ZESs are much less suppressed by temperature. Still, as the temperature is decreased, both the trivial and topological signatures appear much more clearly, reaching the same order of magnitude as their zero-temperature values for temperatures of $\sim 20$~mK. These are quite low temperatures, but are clearly within the range of low temperatures achieved in similar setups in recent experiments \cite{doi:10.1126/science.abf1513}. Combined with the fact that we use realistic parameters throughout this work, we thus believe our results can be experimentally realized.

Lastly, in Fig.~\ref{fig:temperature}(c,d) we show how the topological features still remain enhanced at finite temperatures by using longer S regions, here data taken at $T=20$~mK. This means that, for a fixed temperature, longer S regions can be chosen to improve signal visibility. In Fig.~\ref{fig:temperature} we also observe the oscillations in $S_\pi$ in the topological regime discussed in Section~\ref{section4}. However, they are suppressed and, in the long junction case, distorted, which might make the period doubling achieved by lowering the transparency more difficult to observe.

\subsection{Disorder effects}
As a final note, we also briefly discuss the effect of  disorder, very likely present in  hybrid semiconductor-superconductor platforms with Rashba SOC such as the one studied here. In particular, disorder is expected from scalar impurities or charge inhomogeneities \cite{PhysRevResearch.2.013377,Woods2021, Ahn2021,Zhang2021large,Nichele2017Scaling,aghaee2022inas1} and the question is if it will destroy the distinct signatures in the zero-frequency supercurrent susceptibility. We have verified that the main findings presented in this work remain robust against weak-to-moderately strong Anderson disorder, modeled by random site-dependent fluctuations in the chemical potentials of the N and S regions \cite{cayao2018andreev,Awoga22}.  However, strong disorder softens the topological features in the current susceptibility. Furthermore, we stress that more elaborate disorder models, e.g., including screening effects  \cite{Woods2021, Ahn2021}, may be interesting to pursue in order to provide a realistic description of semiconductor-superconductor systems. For this reason, a detail account of disorder effects on the supercurrent susceptibility will be pursued elsewhere.

\section{Conclusions}
\label{sectionx}

To conclude, we proposed a way to differentiate between trivial and topological zero-energy states in Josephson junctions based on nanowires with Rashba spin-orbit coupling using experimentally accessible phase-biased Josephson measurements. In particular, we explored how the supercurrent and, especially, its zero-frequency supercurrent susceptibility behave as a function of Zeeman field and phase difference across the junction. In particular, we found that the sawtooth profile of the supercurrent in the topological regime, due to the topological MBSs, can be exploited as it produces a discontinuity at $\phi=\pi$, which generates a clear peak in the zero-frequency supercurrent susceptibility at $\phi=\pi$, $S_\pi$. This peak is also not sensitive to the chemical potential in the N region $\mu_N$ as it is a property of the S region. On the other hand, any trivial ZESs in the junction have their origin in the N region and we found that they produce signatures in $S_\pi$ that are inherently sensitive to $\mu_N$. As a consequence, clear peak features in the $S_\pi$ profile as a function of Zeeman field $B$ can be clearly assigned to trivial and topological ZESs by just sweeping the gating in the N region. This result holds for both short and long junctions, despite otherwise quite different properties between them.

We additionally showed that the tuning of the junction transparency can be used to further distinguish between the topological and trivial phases by means of a reduction in the number of MBSs oscillations in $S_\pi$ as transmission is decreased. This happens when the oscillations seen in $S_\pi$ in the topological regime have their period effectively increased, doubled or more, as the MBSs pairs start to oscillate in-phase as the tunneling regime is achieved in the junction. We observed this phenomenon for both short and long junctions, although the latter requires a lower transparency in order to enter the appropriate regime. To support the robustness of our findings, we verified that they hold at low finite temperatures  and for weak-to-moderate disorder, thus making our proposal relevant for experimental implementation.   Finally, we point out that Josephson junctions similar to those studied here have already been fabricated and phase-biased
supercurrents have even been measured
\cite{Nilsson2012,Tiira2017,PhysRevLett.124.226801,PhysRevLett.125.116803,PhysRevLett.126.036802}, which highlights the experimental relevance of the supercurrent
susceptibility \cite{PhysRevLett.122.076802, PhysRevLett.110.217001,
Dassonneville_2018, PhysRevB.97.184505, PhysRevB.88.174505,
PhysRevB.97.041415,PhysRevB.104.L100506} as an additional tool to
distinguish trivial and topological states in superconductor-semiconductor systems.

\section*{Acknowledgments}
	We acknowledge financial support from the Swedish Research Council (Vetenskapsr\aa det Grants No.~2018-03488 and 2021-04121), the G\"{o}ran Gustafsson Foundation (Grant No.~2216), and the Knut and Alice Wallenberg Foundation through the Wallenberg Academy Fellows program. LB acknowledges partial support from the Coordena\c{c}\~ao de Aperfei\c{c}oamento de Pessoal 
	de N\'{\i}vel Superior - Brasil (CAPES) - Finance Code 001. LGDS acknowledges financial support from Brazilian agencies FAPESP (Grant 2016/18495-4) and CNPq (Grants 423137/2018-2, and 309789/2020-6).

\vspace{10pt}

\bibliographystyle{unsrt.bst}
\bibliography{biblio}
\end{document}